\documentclass[iop]{emulateapj}
\usepackage{graphicx}

\newcommand{\eg}{e.g.,\ }

\newcommand{\etal}{et~al.\ }
\newcommand{\ltsima}{$\; \buildrel < \over \sim \;$}
\newcommand{\simlt}{\lower.5ex\hbox{\ltsima}}
\newcommand{\gtsima}{$\; \buildrel > \over \sim \;$}
\newcommand{\simgt}{\lower.5ex\hbox{\gtsima}}
\newcommand{\kms}{km s$^{-1}$}
\newcommand{\magsec}{mag arcsec$^{-2}$}

\newcommand{\deli}{$\Delta i$\ }

\def\muc#1{$\mu_{0,#1}$}
\def\bmv{$B-V$}

\def\re{$r_{\rm e}$}

\def\Lsun{L$_\sun$\ }
\def\magsec{mag arcsec$^{-2}$}

\begin{document}

\title{Galaxies at the extremes: Ultra-diffuse galaxies in the Virgo Cluster}

\author{J. Christopher Mihos,\altaffilmark{1} 
   Patrick R. Durrell,\altaffilmark{2}
   Laura Ferrarese,\altaffilmark{3}
   John J. Feldmeier,\altaffilmark{2}\break
   Patrick C\^ot\'e,\altaffilmark{3}
   Eric W. Peng,\altaffilmark{4,5}
   Paul Harding,\altaffilmark{1}
   Chengze Liu,\altaffilmark{6,7}\break
   Stephen Gwyn,\altaffilmark{3}
   and
   Jean-Charles Cuillandre\altaffilmark{8}
}   

\altaffiltext{1}{Department of Astronomy, Case Western Reserve University,
10900 Euclid Ave, Cleveland, OH 44106, USA}

\altaffiltext{2}{Department of Physics and Astronomy, Youngstown
State University, Youngstown, OH 44555, USA}

\altaffiltext{3}{Herzberg Institute of Astrophysics, National Research
Council of Canada, Victoria, BC V9E 2E7, Canada}

\altaffiltext{4}{Department of Astronomy, Peking University, 
Beijing 100871, China}

\altaffiltext{5}{Kavli Institute for Astronomy and Astrophysics, 
Peking University, Beijing 100871, China }

\altaffiltext{6}{Center for Astronomy and Astrophysics, Department of
Physics and Astronomy, Shanghai Jiao Tong University, Shanghai
200240, China}

\altaffiltext{7}{Shanghai Key Lab for Particle Physics and Cosmology,
Shanghai Jiao Tong University, Shanghai 200240, China}

\altaffiltext{8}{Canada-France-Hawaii Telescope Corporation, Kamuela, HI
96743, USA}

\begin{abstract}

We report the discovery of three large ($R_{29} \gtrsim $ 1\arcmin)
extremely low surface brightness ($\mu_{V,0} \approx 27.0$) galaxies
identified using our deep, wide-field imaging of the Virgo Cluster from
the Burrell Schmidt telescope. Complementary data from the {\it Next
Generation Virgo Cluster Survey} do not resolve red giant branch stars
in these objects down to $i=24$, yielding a lower distance limit of 2.5
Mpc. At the Virgo distance, these objects have half-light radii 3--10
kpc and luminosities L$_{\rm V}=2-9\times 10^7$~\Lsun\llap. These
galaxies are comparable in size but lower in surface brightness
than the large ultradiffuse LSB galaxies recently identified in the Coma
cluster, and are located well within Virgo's virial radius; two are
projected directly on the cluster core. One object appears to be a
nucleated LSB in the process of being tidally stripped to form a new
Virgo ultracompact dwarf galaxy. The others show no sign of tidal
disruption, despite the fact that such objects should be most vulnerable
to tidal destruction in the cluster environment. The relative proximity
of Virgo makes these objects amenable to detailed studies of their
structural properties and resolved stellar populations. They thus
provide an important new window onto the connection between cluster
environment and galaxy evolution at the extremes.

\end{abstract}


\section{Introduction}

Our view of galaxy populations in the universe continues to be shaped by
observational selection effects, particularly that of limiting surface
brightness (Disney 1976). In the field, low surface brightness (LSB)
galaxies exist in significant numbers (\eg McGaugh 1996), but were
largely unnoticed until the CCD imaging surveys of the 1980s (see, \eg
Bothun \etal 1997). Recent observations have probed even
deeper, to limiting surface brightnesses of $\mu_B \approx 28-30$
\magsec, revealing galaxies of ever lower surface brightness. This was
demonstrated most recently through deep wide-field imaging of the Coma
Cluster (van Dokkum \etal 2015a; vD15a, Koda \etal 2015; K15), which
found a population of large (\re=$2-5$ kpc; i.e., Milky Way-sized)
and extremely diffuse galaxies with central surface brightnesses
\muc{V}$=24-26$. These objects preferentially (but not exclusively)
populate the cluster outskirts, with follow-up spectroscopy confirming
that at least one is located within Coma (van Dokkum \etal 2015b;
vD15b).

That such diffuse galaxies exist in a rich cluster like Coma is a
surprise. Because of their low densities and shallow potential wells,
LSB galaxies should be most vulnerable to tidal perturbations as they
move through the cluster, making their lifetimes very short (\eg Moore
\etal 1996). Repeated encounters with other galaxies and with the
cluster potential can whittle away stars from these objects, feeding the
diffuse intracluster light. Complete tidal disruption may leave behind
their high density nuclei, leading to the formation of ultracompact
dwarf galaxies (UCDs; Bekki \etal 2003; Pfeffer \& Baumgardt 2013). How
the large, diffuse galaxies in Coma can survive this dynamically harsh
environment is unclear, suggesting either that they may be falling into
the cluster for the first time, or extremely dark matter dominated
systems and thus more robust against tidal perturbations.

Finding such systems in the nearby Virgo Cluster would be of particular
interest, since they would be close enough ($d_{\rm Virgo}=16.5$ Mpc;
Mei \etal 2007, Blakeslee \etal 2009) to resolve their stellar
populations and study their structure in detail.  Large LSBs in Virgo were
first hinted at in photographic catalogs by Sandage \& Binggeli (1984);
deeper studies subsequently identified diffuse objects with
\muc{V}$=24-26$ and sizes \re$>$ 1 kpc (Impey \etal 1988; Caldwell
2006), somewhat less extreme than the Coma objects. New deep imaging of
Virgo by Mihos \etal (2005 and in preparation), as well as the {\it Next
Generation Virgo Cluster Survey} (NGVS; Ferrarese \etal 2012), now allow
us to search for systems at even lower surface brightness. Here we
report the discovery of three large, extremely diffuse LSBs found in our
deep Virgo imaging. With central surface brightnesses of \muc{V}$\approx
27.0$ and sizes of \re$=3-10$ kpc, these objects are comparable in size
to the vD15a Coma objects, but even lower in surface brightness. All
three objects are located in the inner 0.5 Mpc of Virgo, well within the
virial radius ($R_{vir}=1.55$ Mpc; McLaughlin 1999). With such extremely
low surface brightnesses, and projected deep within the cluster
potential, these objects give us an opportunity for up-close study of
the lowest density galaxies found in the high density cluster
environment.

\begin{figure*}[]
\centerline{\includegraphics[width=7.0in]{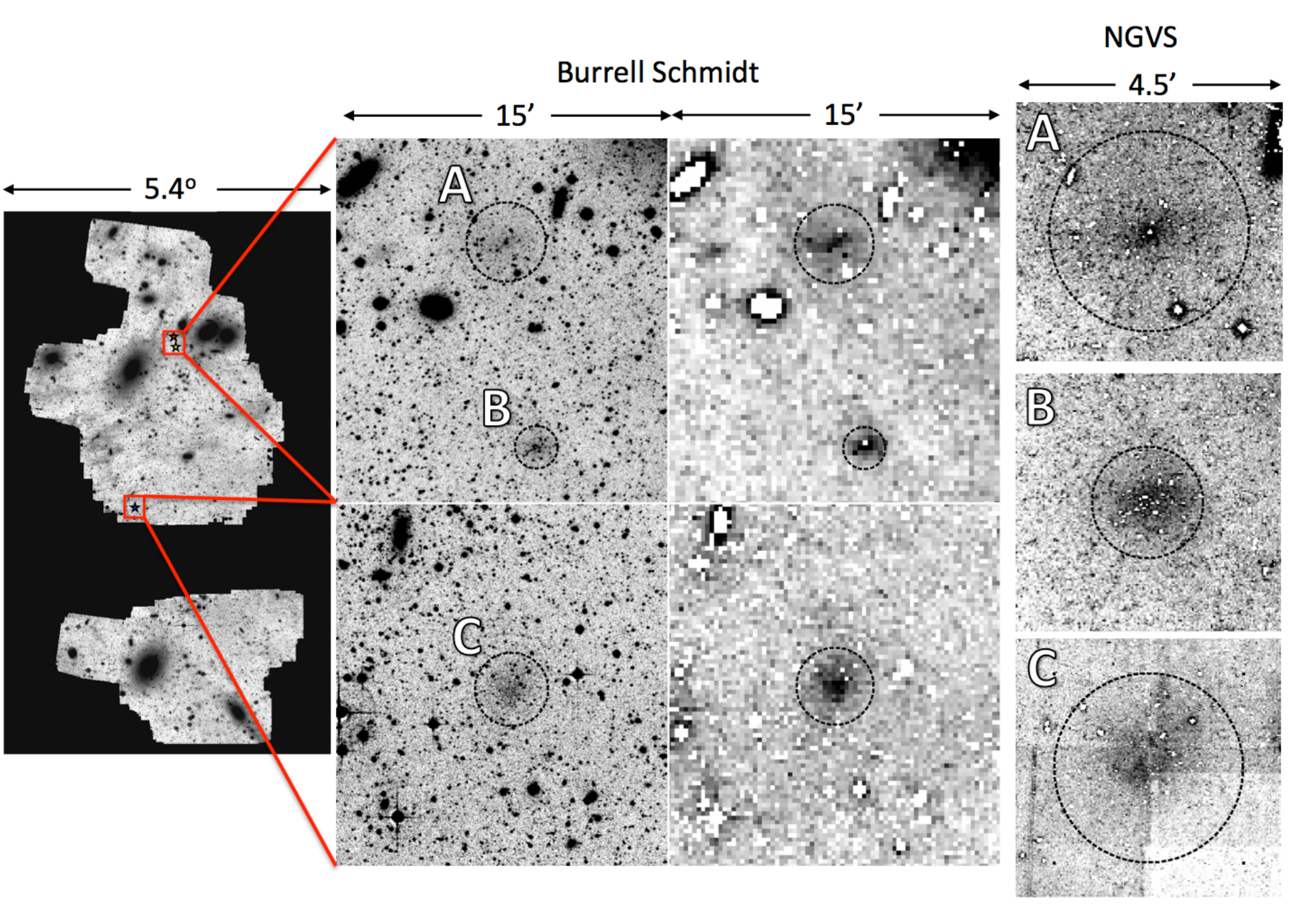}}
\caption{Optical imaging of the Virgo LSBs, with north up and east to
the left. The leftmost panel shows objects' locations within
 the full footprint of the Burrell
Schmidt imaging survey, while the middle panels show the Schmidt imaging
both at full resolution, and masked and rebinned to show faint
structure. The rightmost panels show the NGVS imaging smoothed to
1\arcsec\ resolution.
} 
\label{imaging}
\end{figure*}

\section{Deep Surface Photometry}

To search for ultra diffuse galaxies in Virgo, we use our deep Virgo
imaging survey from CWRU's Burrell Schmidt telescope (Mihos \etal 2005
and in preparation). This survey covers 15.1 (16.3) degree$^2$ down to a
per-pixel limiting surface brightness of 29.0 (28.5) \magsec\ in $B$
($V$) with a pixel scale of 1.45\arcsec\ pixel$^{-1}$. Throughout the
imaging a myriad of small LSB objects can be seen; however, our interest
here was to find the most extreme objects with isophotal sizes of
R$_{\rm V,29} \gtrsim$ 1\arcmin\ and central surface brightnesses
$\mu_{V,0} \gtrsim 26.5$. Because objects at such low surface brightness
and large angular size typically contain many compact, high surface
brightness contaminants (foreground stars and background galaxies),
automated detection algorithms are extraordinarily difficult to employ,
and often miss true objects while making false detections due to
instrumental noise, scattered light, or diffuse galactic cirrus. Rather
than using automated detection, two of us (J.C.M. \& J.J.F.) each made visual searches of the imaging, 
independently identifying three of these
extreme LSBs --- two in the Virgo cluster core, approximately halfway
between M87 and M86, and a third 2\degr\ south of M87, towards the M49
subcluster. Subsequent inspection of the deep NGVS imaging (Ferrarese
\etal 2012) confirms all objects, providing deep $u^*giz$
data\footnote{All magnitudes have been corrected for foreground
extinction (Schlegel \etal 1998), and NGVS magnitudes are given in the
CFHT MegaCam system.} with sub-arcsecond seeing.

Figure~\ref{imaging} shows the deep Burrell Schmidt and NGVS imaging of
our extreme Virgo LSBs. The right panel shows the location of the
objects within Virgo, taken from from Mihos \etal in preparation. The
middle panels show the Schmidt imaging both at full resolution, and
after masking discrete sources and median filtering in 9x9 pixel
(13\arcsec x13\arcsec) scales. The rightmost panels show expanded views
of the NGVS imaging, similarly processed to 1\arcsec\ pixel$^{-1}$
scales. The objects are clearly visible in both datasets. While this
paper focuses on the largest LSB objects, many more smaller objects
exist throughout Virgo; one example can be seen in the middle panels
directly east (left) of VLSB-A: a small LSB galaxy originally identified
in NGVS imaging (Figure~15 of Ferrarese \etal 2012).

\begin{figure*}[]
\centerline{\includegraphics[width=7.0truein]{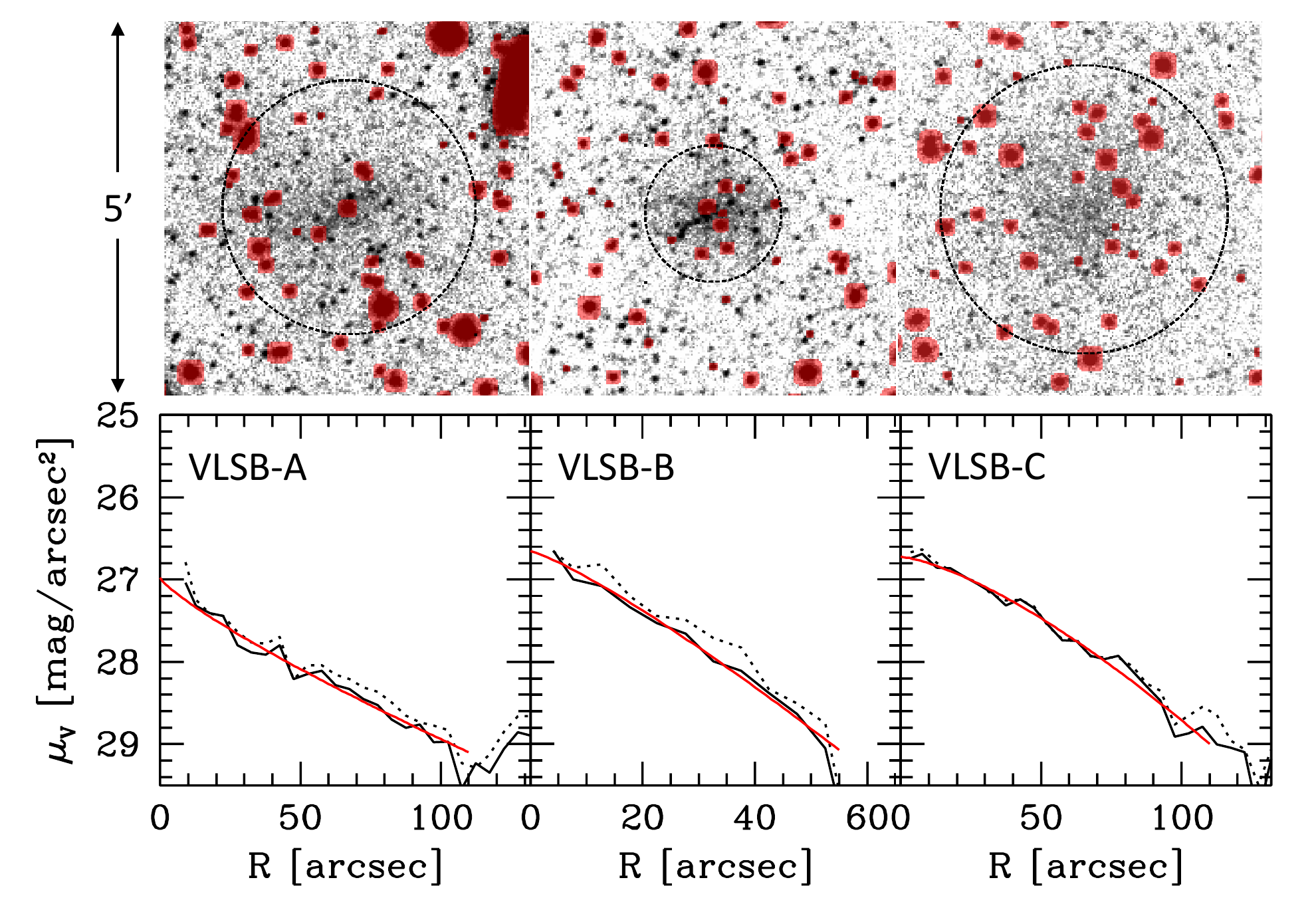}}
\caption{Surface photometry for the Virgo LSBs. 
  Top: Schmidt imaging with photometry masks shown in red. Dashed
  circles show R$_{29}$ for each galaxy. Bottom: Black curves show the
  average (dotted) and median (solid) surface brightness profiles; red curve
  shows Sersic fits to the median profiles. 
}
\label{surfphot}
\end{figure*}

To extract surface brightness profiles of these sources, we first use
IRAF's {\tt objmasks} task to mask the brightest of the compact sources
in and around each object. The sources are likely background
contaminants; NGVS imaging resolves many of them into background
galaxies, and we detect no excess of compact sources over the
surrounding field in either VLSB-A or -C (VLSB-B does appear to have a
slight excess, discussed in \S 3). We then calculate surface brightness
profiles using both the average and median pixel intensities as a
function of radius (Figure~\ref{surfphot}). Using an average includes the
light from the fainter unmasked sources, while a median essentially
traces the diffuse light alone. We fit Sersic models to the median
surface brightness profiles; the fits are shown in Figure~\ref{surfphot},
and the derived structural parameters given in Table~\ref{galprops}.
Ellipticity estimates for VLSB-B and -C come from GALFIT modeling;
VLSB-A is too low in surface brightness to yield a meaningful estimate.
We report average \bmv~colors for VLSB-A and -B, but cannot measure a color for
VLSB-C as it falls outside the $B$ imaging footprint of the Schmidt
survey.

The derived properties given in Table~\ref{galprops} show that, like the extreme
Coma LSBs (vD15a, Koda \etal 2015; K15), these objects are reasonably well
characterized by exponential profiles ($n\sim1.0$). They are large (\re
=40\arcsec--120\arcsec, or 3-10 kpc at the Virgo distance),
extremely low in central surface brightness ($\mu_{V,0} \gtrsim 26.5$),
and fairly red (\bmv=0.6--0.7). A search for the galaxies in other
datasets finds they are undetected both in deep 21-cm ALFALFA data
(Haynes \etal 2011) and far-UV GALEX imaging; coupled with their diffuse
nature and lack of strong H$\alpha$ emission (A. Watkins, private
communication), this suggests they are not actively forming stars.

\begin{figure*}[]
\centerline{\includegraphics[width=6.0truein]{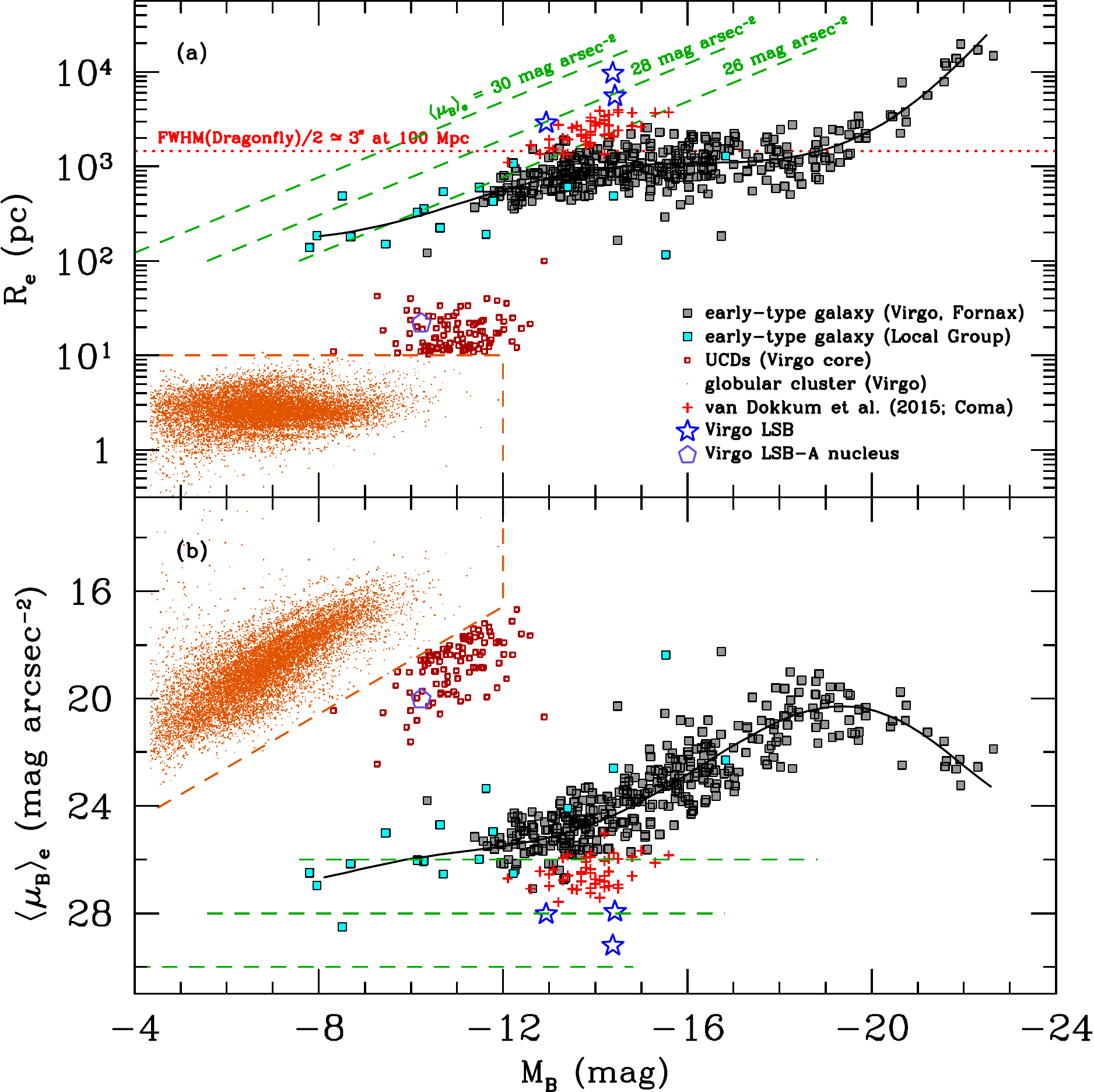}}
\caption{
Structural properties of the Virgo LSBs compared with other stellar
systems, including early type galaxies in the Virgo and Fornax Clusters
and in the Local Group, as well as GCs and UCDs in Virgo, and the
extreme LSBs found in Coma. The dashed orange lines show the GC
selection box, while lines of constant surface brightness are shown in
green. 
}
\label{scaling}
\end{figure*}

Figure~\ref{scaling} compares our VLSB objects to other stellar
systems, including early type galaxies in the Virgo and Fornax clusters
and the Local Group, as well as globular clusters (GCs) and UCDs
in Virgo  (see Ferrarese \etal 2012 for details of the data compilation).
We also include the extreme LSB galaxies in Coma from vD15a.
In this figure, the Coma LSBs merge smoothly onto the
sequence for high surface brightness galaxies, while the extreme Virgo
LSBs reported here populate the plot at the low surface brightness end
(see also K15). In this magnitude range ($M_B=-12$ to $-15$),
early-type cluster galaxies show a continuum in surface brightness,
similar to the distribution of surface brightness in the field (McGaugh
1996). The lack of small (\re\ $\lesssim$ 1 kpc) LSB galaxies in this
plot is a selection effect; in Coma, they fall below the spatial
resolution of vD15a, while in Virgo our selection focused on large
galaxies ($R_{29} \gtrsim$ 1\arcmin). At $\langle \mu_B \rangle_e > 26$,
smaller objects certainly exist; they are seen in the K15 Coma sample
and likely make up a significant fraction of the small LSB objects
visible in Virgo in the Schmidt and NGVS surveys. It's clear from these
studies that LSB galaxies are well-represented in cluster galaxy
populations, once selection effects are factored in.

\section{Discrete Source Imaging}

The preceding discussion presumes the VLSB objects are in fact Virgo
Cluster members. To rule out a more local distance, we
used NGVS imaging to search for any resolved stellar populations
(likely luminous RGB stars). We use the NGVS master catalog to extract
deep $g$ and $i$ photometry for all sources located within each object's R$_{29}$
isophotal radius. Point sources were selected as outlined in Durrell \etal (2014): by defining a
concentration index \deli = $i_{4} - i_{8}$, the difference between the
4-pixel and 8-pixel diameter $i$ aperture-corrected magnitudes, and
identifying point sources as objects with $|$\deli$| < 0.10$. Only the $i$
data was used to define point sources, as the NGVS $i$ images
were taken under the best  seeing conditions.

\begin{figure*}[]
\centerline{\includegraphics[width=7.0truein]{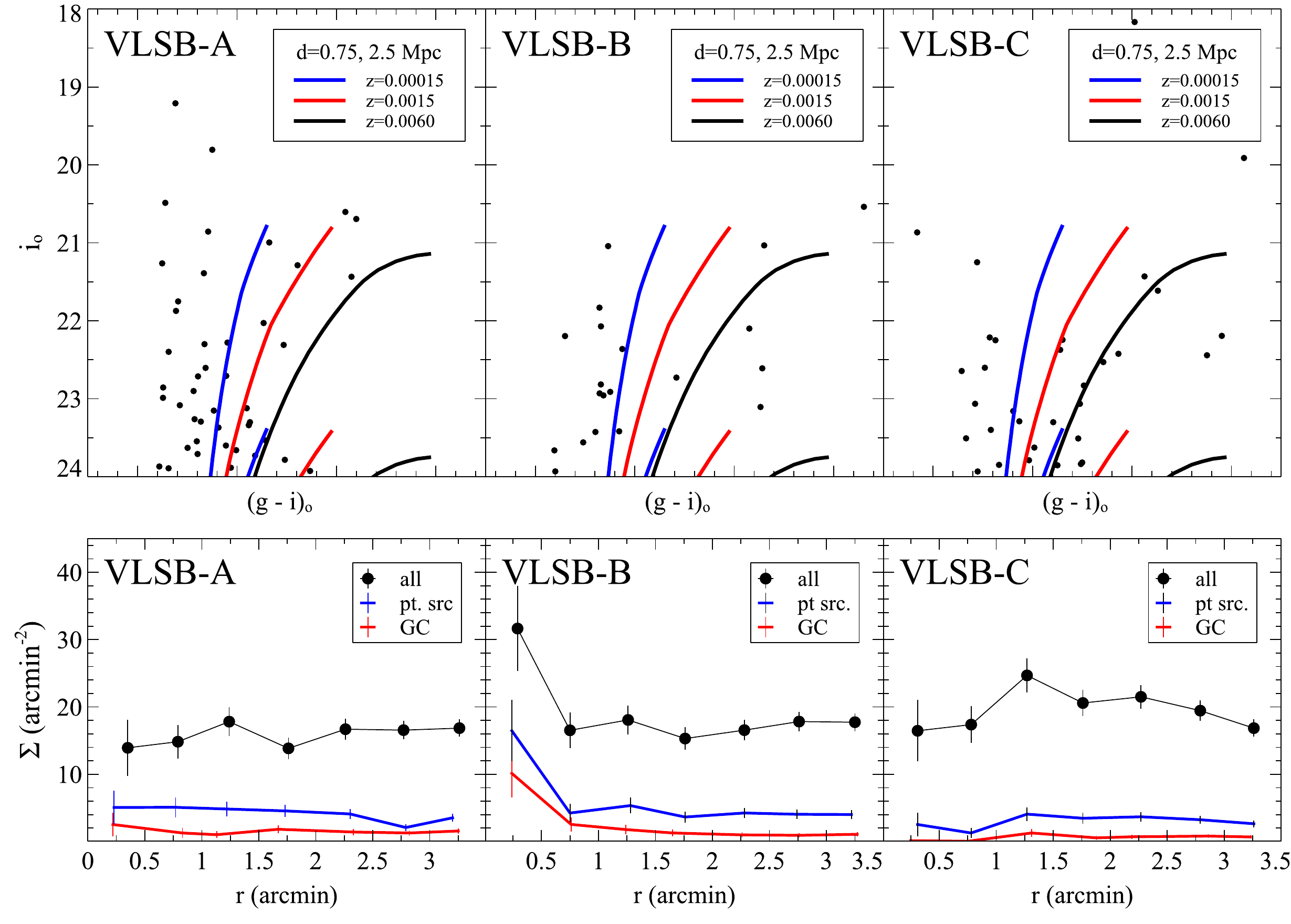}}
\caption{
Photometry of discrete sources from NGVS imaging. Top: CMDs for point
sources within each object, with RGB isochrones at d=0.75 and 2.5 Mpc overplotted
for different metallicities. Bottom: radial density profiles for all discrete sources (black), point
sources (blue), and sources photometrically consistent with Virgo globular clusters (red).
}
\label{ngvs}
\end{figure*}

In Figure~\ref{ngvs} we plot $i, (g-i)$ color-magnitude diagrams (CMDs)
for all point sources with $18<i<24$ within each object. The faint
magnitude cutoff lies $\sim 1$ mag above the S/N$=10$ limit for NGVS $i$
sources, while the bright limit is avoids saturated objects. We also
overplot the PARSEC isochrones of Bressan \etal (2012) with
metallicities $Z=0.00015$, $Z=0.0015$ and $Z=0.0060$ ([M/H]$=-2.0$,
$-1.0$, and $-0.4$, respectively). We plot isochrones for a pair of
distances: $d=0.75$ Mpc, a representative Local Group distance, and
$d=2.5$ Mpc, the largest distance for which we could detect the most
luminous RGB stars.

From the CMDs, we see no clear detection of any RGB population. For each
isochrone distance, we can estimate the expected number of RGB stars by
scaling the RGB population detected in a low luminosity ($M_V=-10.6$)
Virgo dSph galaxy (Durrell \etal 2007). At 0.75 Mpc, we would expect 54,
21, and 96 RGB stars down to $i=23.5$ in VLSB-A, -B, and -C
respectively. At 2.5 Mpc, the expected RGB counts become 38, 15, and 66.
Comparing these numbers to the paucity of the stellar sources along the
RGB tracks shown in Figure~\ref{ngvs}, we conclude that all three of
these objects must lie beyond 2.5~Mpc.

\begin{deluxetable}{cccc}
\tabletypesize{\scriptsize}
\tablewidth{0pt}
\tablecaption{Extreme LSB Galaxies in Virgo\tablenotemark{a}}
\tablehead{\colhead{ } & \colhead{VLSB-A\tablenotemark{b}} & \colhead{VLSB-B} & \colhead{VLSB-C}}
\startdata
RA & \ 12:28:15.9 & \ 12:28:10.6 & \ 12:30:37.3 \\
Dec & +12:52:13 & +12:43:28 & +10:20:53 \\
R$_{29}$ & 103\arcsec (26\arcsec) & 55\arcsec (4\arcsec)  & 110\arcsec (5\arcsec) \\
$\mu_{V,0}$ & 27.0 (0.30) & 26.7 (0.11) & 26.7 (0.08)\\ 
$\langle\mu\rangle_{e,V}$ & 28.5 (0.30) & 27.5 (0.11) & 27.4 (0.08)\\ 
\re & 121\arcsec (24\arcsec) & 36\arcsec (2\arcsec) & 69\arcsec (3\arcsec)\\
Sersic $n$ & 1.2 (0.22) & 0.8 (0.08) & 0.7 (0.05) \\
$m_V$ & 16.1 (0.53) & 17.6 (0.16) & 16.2 (0.11) \\
$(B-V)$ & 0.7 (0.1) & 0.6 (0.1) & --- \\
Ellipticity& --- & 0.17 (0.15) & 0.12 (0.15) \\
M$_V$\tablenotemark{c}  & $-$15.0 & $-$13.5 & $-$14.9 \\
\re\tablenotemark{c}  & 9.7 kpc & 2.9 kpc & 5.5 kpc 
\enddata
\tablenotetext{a}{parameter uncertainties are shown in parentheses.}
\tablenotetext{b}{photometric properties for VLSB-A do not include its
compact nucleus.
}
\tablenotetext{c}{adopting $d_{\rm Virgo}$=16.5 Mpc.}
\label{galprops}
\end{deluxetable}

As a further check, Figure~\ref{ngvs} also plots the radial number
density profile of discrete sources (both resolved and point-like)
within 3.5\arcmin\ of each object. If the VLSB objects are nearby, at
$d<$2.5 Mpc, their resolved stellar populations should appear as an
increased density of point sources near the centers of each object. At
Virgo-like distances, an excess of discrete sources near the galaxies
would trace star clusters rather than individual stars, while a lack of
concentration would argue the sources are contaminants (predominantly
background galaxies or foreground MW stars) unassociated with the VLSB
objects entirely.

Figure~\ref{ngvs} shows no excess of of either point sources or resolved
sources associated with VLSB-A or -C --- the discrete sources are
consistent with pure background contamination.
Thus, VLSB-A and -C are indeed
diffuse, with no sign of either resolved stars or star clusters. VLSB-B
presents an interesting contrast, however, as we find an excess of both
resolved sources and point sources within the galaxy. Several of these
sources appear to be background galaxies, and the point source excess
appears not to be resolved RGB stars, but rather a small population of
globular clusters. We show this by also plotting in Figure~\ref{ngvs}
the density of sources with properties expected for Virgo globular clusters, 
using the selection criteria of Durrell \etal (2014): stellar or
only slightly resolved sources ($-0.10<$ \deli $< 0.15$) with
$0.55<g-i<1.15$ and $19.5<i<23.5$. We see a modest excess of these
candidate globulars in VLSB-B, with $N=6\pm 3$ objects in the
central 0.5\arcmin, after removal of elongated objects (presumably
background galaxies) and subtraction of a local background. This
tentative detection of globular clusters supports the conclusion that
VLSB-B is an extremely diffuse LSB galaxy located within Virgo.

\section{Notes on Individual Objects}

{\bf VirgoLSB-A} is projected deep within the Virgo core, 0.75\degr\ NW
of M87 and 0.5\degr\ ESE of M86. It appears as a nucleated LSB galaxy
with long, arcing tidal stream that runs NE-SW through the
galaxy and extending off the frame in Figure~\ref{imaging} (the full
extent of the stream can be seen in Figure~1 of Mihos \etal 2005). The
tidal stream is curved concave to M86 (and to the nearby galaxy pair NGC
4435/8), suggesting VLSB-A may be orbiting within the M86 subgroup rather
than around M87 itself. The LSB component of the
system is quite extended
and shows a bar-like
component oriented at 135\degr.

VLSB-A's nucleus is marginally resolved in our NGVS imaging; a structural analysis
of the nucleus using KINGPHOT (Jord\'an \etal 2005, Liu \etal 2015) yields an effective radius of
$r_{e,g}=0.27$\arcsec\ (22~pc). The nucleus has a radial velocity of $-120
\pm 40$ \kms\ (Peng \etal in preparation), quite distinct from M87
($+1064$ \kms) and offset by $\sim 2\sigma_v$ from the mean velocity of
Virgo E/S0 galaxies ($\langle v \rangle = 1017$ \kms, $\sigma_v=589$
\kms; Binggeli \etal 1993). However, its similarity in velocity to M86
($-224$ \kms), again argues that VLSB-A is part of Virgo's M86 subgroup.

On the whole, the properties of VLSB-A clearly suggest we are witnessing
the dynamical formation of a new cluster UCD, made via tidal threshing
of a low mass cluster galaxy (\eg Bekki \etal 2003; Pfeffer \& Baumgardt
2013). The tidal stream and bar-like morphology of the galaxy are
consistent with the response of a low mass galaxy to a strong tidal
field, while the galaxy's red \bmv\ color suggests that star formation
in the system has ceased. Figure~\ref{scaling} compares the structural
properties of the nucleus of VLSB-A to UCDs in the Virgo core (Zhang
\etal 2015, Liu \etal 2015), where it can be seen that the nucleus lies
in the large, low surface brightness tail of the UCD distribution.
VLSB-A is likely in a short-lived transitory phase, as the cluster
environment strips its diffuse outskirts to form a new Virgo UCD.

{\bf VirgoLSB-B} is also projected onto the cluster core, only 9\arcmin\
south of VLSB-A. However, unlike VLSB-A it shows a more regular
morphology with no obvious tidal debris, and is somewhat bluer (\bmv=0.6)
as well. As noted previously, this object shows a significant population
of sources 
photometrically consistent with Virgo
globular clusters. Adopting a Gaussian globular
cluster luminosity function with a turnover at $g_{\rm TO}=23.8 \pm 0.2$
and $\sigma=1$ mag (Jord\'an \etal 2007), the total inferred GC
population is $N_{\rm GC,tot}=9 \pm 4.5$. This yields a specific
frequency of $S_N = 40 \pm 20$, rather large (albeit with large
uncertainties) for galaxies of this luminosity, which typically have
$S_N = 10-20$ with large spread (Peng \etal 2008, Georgiev \etal
2010).

Finally, {\bf VirgoLSB-C} is found 2\degr\ (575 kpc or $\sim {1\over 3}
R_{vir}$) south of M87, between the Virgo A and M49 subclusters. The object appears to be
purely diffuse, with no excess of compact sources over background in the
system, and shows no obvious sign of tidal stripping.

\section{Discussion}

The three objects presented here are quite diverse in their physical
properties. While they are all large and extremely diffuse, and
projected deep within Virgo, only one (VLSB-A) shows obvious signs of
the tidal damage expected for diffuse galaxies in a dense environment;
the other two are quite round ($\epsilon < 0.2$) with no morphological
deformation or extended tidal debris. Meanwhile, globular clusters are
only detected within VLSB-B, which suggests a surprisingly high specific
frequency; neither VLSB-A or -C show evidence for globular clusters,
yielding upper limits of $S_N$ \simlt $2-3$ for these objects.

The differences between the objects may be due to differences in their
evolutionary state or local environment. The tidal morphology of VLSB-A,
along with its kinematic association with M86, strongly argues that the
object is interacting within Virgo's M86 subgroup. However, the lack of
obvious tidal distortion in VLSB-B and -C, diffuse galaxies which should
be most vulnerable to cluster tides, suggests instead they may lie in
the cluster outskirts, or be falling into Virgo for the first time.
Alternatively, they may be very dark matter dominated, like field LSBs
(\eg de Blok \& McGaugh 1997), and therefore more resistant to tidal
stripping. In this context, it is interesting that VLSB-B shows evidence
for globular clusters. If globular cluster populations trace the dark
matter content of a galaxy (\eg Blakeslee \etal 1997; Peng \etal 2008;
Harris \etal 2013; Hudson \etal 2014), the high specific frequency we
infer for VLSB-B may be a signature of a massive dark halo that protects
the system from rapid tidal destruction. However, a detailed
understanding of how these objects fit into the picture of
dynamically-driven galaxy evolution in clusters demands a better
determination of both their local environment and their intrinsic
properties.

While the properties of VLSB-A convincingly place it deep within Virgo,
the situation for VLSB-B and -C is less clear. Without direct distance
estimates (such as from the TRGB), our results do not unambiguously
locate the objects within Virgo. While we rule out a Local Group
distance, they may lie in the field along the line of sight, either in
front of or beyond Virgo. However, arguments that these are field
objects merely projected onto the Virgo Cluster also run into problems,
since their gas-poor nature and lack of knotty structure makes them very
different from known field LSBs (\eg de Blok \& McGaugh 1997).
Furthermore, if the objects lie beyond Virgo, their physical properties
would be even more extreme --- larger and more luminous (at fixed low
surface brightness), potentially rivaling giant LSB galaxies such as
Malin~1. Even if located in the intervening field, at distances 2.5--15
Mpc, their large angular sizes and low surface brightnesses show they
still inhabit regions of structural parameter space that have been
largely unexplored.

Nonetheless, the properties of VLSB-A, and the projection of all three
galaxies well within Virgo's virial radius, argue that they are indeed
associated with the Virgo itself. The presence of extremely diffuse
galaxies in Virgo (this paper) and Coma (vD15ab, K15) shows these
objects populate a range of cluster environments. While ultradiffuse
Coma galaxies have been reported in greater numbers than identified
here, several factors make direct comparison difficult. First, our
Schmidt imaging covers only the inner $\sim$ 15\% of Virgo; correcting
for survey area suggests a total of $\sim$ 20 objects throughout the
cluster, and even more if they are located preferentially in the cluster
outskirts (as found in Coma; vD15a). Second, very few of the reported
Coma objects approach the low surface brightnesses of our VLSB objects
(see Figure~\ref{scaling}); such systems may simply be intrinsically
rare in {\it both} clusters. Finally, Coma is a much richer cluster than
Virgo, and may house more galaxies of {\it all} types, LSBs included.
However, in comparing LSB populations, cluster richness is a
double-edged sword: a richer cluster may also make for a harsher
dynamical environment that shortens their lifetime and reduces their
overall numbers.

Ultimately, these questions will be best addressed through a
cluster-wide census of the Virgo LSB galaxy population using wide-field
surveys such as the NGVS. Identifying a larger sample of ultradiffuse
galaxies in Virgo would also allow for detailed studies of their
resolved stellar populations and physical structure on spatial scales
not possible in more distant clusters. Thus a more systematic search for
this elusive galaxy population in Virgo is well-motivated.

\end{document}